\documentclass[seceq]{ptptex}

\def\Vec#1{\mbox{\boldmath$#1$}}
\usepackage{graphics}

\title{
Canonical Quantization for a Relativistic Neutral Scalar Field \\
in Non-equilibrium Thermo Field Dynamics
}

\author{
Yuichi \textsc{Mizutani}$^1$, Tomohiro \textsc{Inagaki}$^2$, \\
Yusuke \textsc{Nakamura}$^3$ and Yoshiya \textsc{Yamanaka}$^3$ } 

\inst{
  $^1$Department of Physics, Hiroshima University, 
 Higashi-Hiroshima, Hiroshima 739-8526, Japan \\
 $^2$ Information Media Center, Hiroshima University,
 Higashi-Hiroshima, Hiroshima 739-8521, Japan \\
 $^3$ Department of Electronic and Photonic Systems, 
 Waseda University, Tokyo 169-8555, Japan
}

\abst{
A relativistic neutral scalar field is investigated in non-equilibrium 
thermo field dynamics. The canonical quantization is applied to the 
fields out of equilibrium. Because the thermal Bogoliubov transformation
becomes time-dependent, the equations of motion for the ordinary unperturbed 
creation and annihilation operators are modified. This forces us to introduce
a thermal counter term in the interaction Hamiltonian which generates 
additional radiative corrections. 
Imposing the self-consistency renormalization condition on the 
total radiative corrections, we obtain the quantum Boltzmann equation for the 
relativistic scalar field. 
}

\begin{document}

\maketitle

\section{Introduction}
A scattering process is described 
by a relativistic quantum field theory at high energy. 
The quantum field theory has to be extended to 
study an out of equilibrium system for a relativistic field. 
There are several formalisms to 
introduce a thermal property in the quantum field theory, 
for example, thermo field dynamics, closed time path 
formalism and Langevin equation\cite{lngvn1,ctp1,ctp2}. 
The thermo field dynamics (TFD) is a real time formalism 
based on the canonical quantization.\cite{umezawa1,finitetfd1,CQTFD0,netfd2} 
In TFD the thermal Bogoliubov transformation is introduced with 
particle number density and the thermal Bogoliubov transformed 
oscillator defines the so-called thermal vacuum state. 
The thermal average of a dynamical operator 
is represented as an expectation value in the thermal vacuum state. 

Much attention has been paid to generalize it for an out of equilibrium system 
and derive the time evolution equation for a particle number distribution. 
The non-equilibrium thermo field dynamics (NETFD) 
is constructed by extending the thermal Bogoliubov transformation 
in Ref.~\citen{netfd1}. 
The Boltzmann-like equation for a non-relativistic particle 
is found by the time evolution equation for the expectation value of 
the particle number operator with perturbed oscillators 
in Ref.~\citen{matsumoto3}. 

In NETFD the thermal counter term is 
introduced from the consistency of the time evolution for 
an ordinary oscillator and the thermal Bogoliubov transformed one 
\cite{CQNETFD1,ThermC1,CQNETFD2}. 
The Boltzmann equation is derived from 
the self-consistency renormalization condition 
which is imposed on the quantum corrections with the thermal counter
term.\cite{pertN1,pertN2} 
The method was extended to an inhomogeneous system
with diffusion.\cite{inhomo1,inhomo2}
Recently NETFD is applied to evaluate the Bose-Einstein condensation in 
trapped cold atom systems.\cite{ColdAtm1,ColdAtm2,ColdAtm3,ColdAtm4} 

NETFD for a relativistic field is necessary to study the 
thermalization process of quarks and gluons in 
ultrarelativistic heavy ion collisions. 
An out of equilibrium system for a relativistic 
field is also essential in critical phenomena at early universe. 
Several works have been done for a relativistic field in NETFD\cite{rela1,prop1}. 
However, the canonical formalism of NETFD 
has not been fully established for a relativistic field yet. 
Thus we have launched our plan to make a 
systematic study of the canonical quantization in 
NETFD based on the thermal Bogoliubov transformation. 

In this paper we focus on a relativistic 
neutral scalar field and investigate the canonical 
quantization in a homogeneous system. 
In \S 2 we briefly review TFD for a scalar field. 
The thermal Bogoliubov transformation is made time-dependent
for an out of equilibrium system in NETFD, in accordance with 
time-dependent particle number distribution. 
In \S 3 we discuss how to decompose the neutral scalar field. 
In \S 4 the scalar field is quantized in the canonical formalism. 
We introduce the thermal counter term 
and calculate the scalar propagator. 
The self-consistency condition is introduced from the structure of the 
scalar propagator. 
The time evolution equation is obtained from the self-consistency 
renormalization condition. 
In \S 5  it is confirmed that the self-consistency condition implies 
the correspondence between the thermal Bogoliubov parameter 
and the particle number density, given by an expectation of 
the Heisenberg number density operator. 
We impose the condition on the neutral scalar field with self-interactions. 
Concluding remarks are given in \S 6. 

\section{Non-equilibrium thermo field dynamics}
There are several real time formalisms to introduce 
thermal dynamics in a quantum field theory. 
TFD is one of the real time formalisms based on the canonical quantization. 
In TFD the statistical average of an observable quantity 
is represented as an expectation value of the observable 
operator under a pure state called thermal vacuum.
The ordinary oscillator operators, after each degree of freedom is doubled
as below, are related to new oscillator ones through a Bogoliubov transformation
whose annihilation operator annihilates the thermal vacuum. 
The thermal Bogoliubov transformation is made time-dependent to 
deal with a non-equilibrium state in NETFD.
In this section we briefly review the framework of NETFD.

In TFD the thermal degree of freedom is introduced by
doubling the Fock space which is spanned by ordinary annihilation, 
creation operators, $(a\ \  a^\dagger)$, and tilde ones, 
$(\tilde{a}\ \ \tilde{a}^\dagger)$. For a bosonic field the 
operators satisfy the commutation relations, 
\begin{eqnarray}
&& [a_p, a_k^\dagger ]= (2\pi)^3 \delta^{(3)}(\Vec{p}-\Vec{k}),\\
&& [\tilde{a}_p, \tilde{a}_k^\dagger ]
 = (2\pi)^3 \delta^{(3)}(\Vec{p}-\Vec{k}),\\
&& [a_p, \tilde{a}_k^\dagger ]= 0,\ \ 
 [a_p, \tilde{a}_k ] = 0.
\label{cr:a}
\end{eqnarray}
We define the tilde conjugation rules by
\begin{eqnarray}
&& (A_1 A_2)^\sim=\tilde{A_1}\tilde{A_2},  \label{tilde1}\\
&& (c_1 A_1 + c_2 A_2)^\sim=c_1^\ast \tilde{A}_1 + c_2^\ast \tilde{A}_2, \label{tilde2} \\
&& (\tilde{A})^\sim=A,	\label{tilde3}
\end{eqnarray}
where $A_1$ and $A_2$ mean any operators, and $c_1$ and $c_2$ are c-numbers.

Time evolution for the tilde operator is generated by
the tilde conjugate of a usual Hamiltonian.
The tilde-Hamiltonian, $\tilde{H}$, is constructed only from 
the tilde-operators. The time evolution of whole a physical 
system is described by a hat-Hamiltonian,
\begin{eqnarray}
\hat{H} \equiv H-\tilde{H}.
\end{eqnarray}

To develop the canonical quantization in NETFD
it is more convenient to rewrite the annihilation and creation 
operators by the transformation called the time-dependent 
thermal Bogoliubov transformation, 
\begin{eqnarray}
&&\xi_p^\alpha (t) 
= B(n_p(t))^{\alpha \beta} a_p^{\beta} (t), \label{TBT1} \\
&&\bar{\xi}_p^\alpha (t)
= \bar{a}_p^{\beta} (t) B^{-1}(n_p(t))^{\beta \alpha}, \label{TBT2}
\end{eqnarray}
where the upper indices, $\alpha$ and $\beta$, represent the thermal doublets,
\begin{eqnarray}
&& a_p^\alpha = 
\left(
   \begin{array}{c}
	a_p \\
	\tilde{a}_p^\dagger
   \end{array}
\right) ,\ \ \ 
\bar{a}_p^\alpha =
\left(
   \begin{array}{cc}
	a_p^\dagger & -\tilde{a}_p
   \end{array}
\right) , \\
&& \xi_p^\alpha = 
\left(
   \begin{array}{c}
	\xi_p \\
	\tilde{\xi}_p^\dagger
   \end{array}
\right) ,\ \ \ 
\bar{\xi}_p^\alpha =
\left(
   \begin{array}{cc}
	\xi_p^\dagger & -\tilde{\xi}_p
   \end{array}
\right).
\end{eqnarray}
The thermal Bogoliubov matrices, $B$ and $B^{-1}$, are chosen as
\begin{eqnarray}
&& B(n_p(t))=
\left(
   \begin{array}{cc}
	1+n_p(t) & -n_p(t) \\
	-1 & 1 \\
   \end{array}
\right), \label{ns_tbm1} \\ 
&& B^{-1}(n_p(t))=
\left(
   \begin{array}{cc}
	1 & n_p(t) \\
	1 & 1+n_p(t) \\
   \end{array}
\right), \label{ns_tbm2}
\end{eqnarray}
so that the Dyson-Wick formalism can be used.\cite{FynPert1} 
Since we work in the interaction picture, the operators, 
$\xi_p(t)$ and $a_p(t)$, depend on time. 
In a particular case of equilibrium, the Bogoliubov parameter $n_p$
is time-independent and is taken to be the Bose-Einstein distribution. 
The time dependence of the transformed oscillators is given by
\begin{eqnarray}
&&\xi_p^\alpha (t_x)
 = \xi_{p}^\alpha e^{-i \omega_p  t_x}, \label{ns_engy_xi1} \\
&&\bar{\xi}_p^\alpha (t_x)
 =\bar{\xi}_{p}^{\alpha} e^{i \omega_p  t_x}, \label{ns_engy_xi2}
\end{eqnarray}
where $\omega_p$ describes the on-shell energy for the
scalar field, $\omega_p= \sqrt[]{\Vec{p}^2+m^2}$.
Though the energy, $\omega_p$, is not always 
time-independent, we confine ourselves to a system 
with a time-independent $\omega_p$, for simplicity. 
It should be noted that non-trivial time dependence
is induced for the oscillators, $a_p$ and 
$\bar{a}_p$, through the time-dependent Bogoliubov parameter with momentum index,
$n_p(t)$.
For a homogeneous and isotropic system which we assume 
in our practical calculations below, 
$n_p(t)$ is a function of the time and the magnitude of the momentum. 
As will be seen in \S 5, the correspondence between 
$n_p(t)$ and the particle number density, given by an expectation of 
the Heisenberg number density operator, is shown by imposing 
the self-consistency renormalization condition for a 
non-relativistic system \cite{umezawa1,pertN1,pertN2}. 

The thermal vacuum is defined by 
\begin{eqnarray}
&& \xi_p |\theta\rangle = \tilde{\xi}_p |\theta\rangle=0, 
\label{theta1} \\
&& \langle \theta| \xi_p^\dagger = \langle \theta| \tilde{\xi}_p^\dagger = 0.
\label{theta2}
\end{eqnarray}
The thermal vacuum state is invariant under the tilde conjugation, 
$(|\theta \rangle )^{\sim} =|\theta\rangle$, and
$(\langle \theta|)^{\sim} =\langle \theta|$. 
After we perform the thermal Bogoliubov transformations 
(\ref{TBT1}) and (\ref{TBT2}) on the oscillators, $\xi$ 
and $\tilde{\xi}$, Eqs.~(\ref{theta1}) and (\ref{theta2}) 
are rewritten as
\begin{eqnarray}
&& (1+n_{p}(t)) a_p(t) |\theta \rangle = n_{p}(t) \tilde{a}_p^{\dagger}(t)|\theta\rangle, 
\label{ns_tsc_1} \\
&& n_{p}(t) a_p^\dagger (t) |\theta \rangle = (1+n_{p}(t)) \tilde{a}_p(t)|\theta\rangle, 
\label{ns_tsc_2} \\
&& \langle \theta| a_p(t)= \langle \theta| \tilde{a}_p^\dagger(t), 
\label{ns_tsc_3} \\
&& \langle \theta| a_p^\dagger (t)= \langle \theta| \tilde{a}_p(t). 
\label{ns_tsc_4}
\end{eqnarray}
 Note that the thermal bra  and ket vacua are not symmetric due to the choice of the
thermal Bogoliubov matrices in Eqs.~(\ref{ns_tbm1}) and (\ref{ns_tbm2}). 
From Eqs.~(\ref{ns_tsc_3}) and (\ref{ns_tsc_4}) we can
show that the interaction hat-Hamiltonian, $\hat{H}_{int}$, 
satisfies the following condition \cite{umezawa1},
\begin{eqnarray}
\langle \theta|\hat{H}_{int} = 0, \label{HI-C1}
\end{eqnarray}
 (but $\hat{H}_{int}|\theta\rangle \neq 0$),
which enables us to use the Dyson-Wick formalism. 
As is known, the Bogoliubov transformation keeps the commutation relations.
Thus the transformed operators, $\xi$ and $\tilde{\xi}$, satisfy the commutation 
relations. 
\begin{eqnarray}
&& [\xi_p, \xi_k^\dagger ]= (2\pi)^3 \delta^{(3)}(\Vec{p}-\Vec{k}),\\
&& [\tilde{\xi}_p, \tilde{\xi}_k^\dagger ]
 = (2\pi)^3 \delta^{(3)}(\Vec{p}-\Vec{k}),\\
&& [\xi_p, \tilde{\xi}_k^\dagger ]= 0,\ \ 
 [\xi_p, \tilde{\xi}_k ] = 0. 
\end{eqnarray}
A scalar field is quantized under these commutation relations.
It should be noticed that physical observables are constructed by the 
original operators, $a$ and $a^\dagger$. We would like to evaluate the 
expectation value for such operators under the thermal vacuum, 
$|\theta\rangle$.

\section{Relativistic neutral scalar field in NETFD}
In the previous section we have introduced the bosonic operators, $a_{p}$ 
and $\xi_{p}$. A scalar field can be represented by either of 
these operators. Since the thermal Bogoliubov transformation 
does not depend on the time variable in equilibrium, 
the time dependence of both the operators is described by the same
 Hamiltonian in the
interaction picture. On the other hand, the time-dependent 
Bogoliubov transformation induces a discrepancy between the 
time evolution for $a_{p}$ and $\xi_{p}$ in NETFD. 
We introduce a decomposition of the scalar field in terms of 
the operator $a_{p}$ in a consistent manner. 

A neutral scalar field can be represented by the operators, 
$\xi_{p}$ and $\bar{\xi}_{p}$, in the interaction picture,
\begin{eqnarray}
&&  \phi_{\xi}^{\alpha}(x) \equiv \int \frac{d^3\Vec{p}}{(2\pi)^3}
 \frac{1}{\sqrt{2\omega_p}} 
( \xi_{p}^{\alpha}(t_x) 
 e^{ i \Vec{p} \cdot \Vec{x}} 
+ (\tau_3 \bar{\xi}_p (t_x)^T)^\alpha 
 e^{ -i \Vec{p} \cdot \Vec{x}}), \label{phixi1} \\
&&  \bar{\phi}_{\xi}^{\alpha}(x) \equiv \int \frac{d^3\Vec{p}}{(2\pi)^3} 
\frac{1}{\sqrt{2\omega_p}} 
(\bar{\xi}_{p}^{\alpha}(t_x) 
 e^{ -i \Vec{p} \cdot \Vec{x}} 
+ (\xi_p (t_x)^{T} \tau_3)^\alpha 
 e^{ i \Vec{p} \cdot \Vec{x}}), \label{phixi2}
\end{eqnarray}
where $\tau_3$ is the third Pauli matrix acting on thermal indices. 
The canonical conjugates for the fields, $\phi_{\xi}$ and 
$\bar{\phi}_{\xi}$, are given by
\begin{eqnarray}
&&  \pi_{\xi}^\alpha(x) \equiv (-i)\int \frac{d^3\Vec{p}}{(2\pi)^3}
 \sqrt{\frac{\omega_p}{2}} 
( \xi_{p}^\alpha (t_x) 
 e^{ i \Vec{p} \cdot \Vec{x}} 
- (\tau_3 \bar{\xi}_p(t_x)^T)^\alpha 
 e^{ -i \Vec{p} \cdot \Vec{x}}),  \label{pixi1} \\
&&  \bar{\pi}_{\xi}^\alpha(x) \equiv (-i)\int \frac{d^3\Vec{p}}{(2\pi)^3} 
 \sqrt{\frac{\omega_p}{2}} 
( -\bar{\xi}_{p}^\alpha (t_x) 
 e^{ -i \Vec{p} \cdot \Vec{x}} 
+ (\xi_p (t_x)^{T} \tau_3)^\alpha 
 e^{ i \Vec{p} \cdot \Vec{x}}). \label{pixi2}
\end{eqnarray}
These fields satisfy the canonical commutation relations at the equal time, 
\begin{eqnarray}
&& [\phi_{\xi}^\alpha (t,\Vec{x}),\pi_{\xi}^\beta (t,\Vec{y})]
 = i \delta^{(3)}(\Vec{x}-\Vec{y}) \tau_3^{\alpha \beta},\\
&& [ \bar{\phi}_{\xi}^\alpha (t,\Vec{x}),\pi_{\xi}^\beta (t,\Vec{y}) ]
 = i \delta^{(3)}(\Vec{x}-\Vec{y}) \delta^{\alpha \beta}.
\end{eqnarray}
It is straightforward to quantize the scalar fields, 
$\phi_\xi$ and $\bar{\phi}_\xi$, under the thermal vacuum in a canonical 
formalism. However, a canonical formalism is necessary for 
a scalar field written by the operators $a$ and $\bar{a}$ in 
order to evaluate physical observables.

The time dependence of the operators, $a_p$ and $\bar{a}_p$, 
can be fixed by the time evolution equations for the 
operators, $\xi_{p}$ and $\bar{\xi}_{p}$. 
Differentiating Eqs.~(\ref{ns_engy_xi1}) and (\ref{ns_engy_xi2}) 
with respect to the time variable, we obtain
the time evolution equations for the operators, $\xi_{p}$ 
and $\bar{\xi}_{p}$,
\begin{eqnarray}
&& \partial_{t_x} \xi_p^\alpha (t_x)
 = -i \omega_p \xi_p^\alpha (t_x), \label{EOMXi1} \\
&& \partial_{t_x} \bar{\xi}_p^\alpha (t_x)
 = i  \omega_p \bar{\xi}_p^\alpha (t_x). \label{EOMXi2}
\end{eqnarray}
Applying the thermal Bogoliubov transformation (\ref{TBT1})
for the operator $\xi$ in Eq.~(\ref{EOMXi1}), we obtain a
time evolution equation for the operator, $a$,
\begin{eqnarray}
\partial_{t_x}a_p^{\alpha}(t_x)
=-i(\omega_p -i \dot{n}_p(t_x)T_0)^{\alpha \beta} a_p^{\beta}(t_x), \label{EOMa01}
\end{eqnarray}
where the matrix, $T_0$, is  
\begin{eqnarray}
T_0 =
\left(
   \begin{array}{cc}
	1 & -1 \\
	1 & -1 \\
   \end{array}
\right).
\end{eqnarray}
We note that ${T_0}^2 = 0$. From Eq.~(\ref{EOMa01}) it is 
found that the energy eigenvalue for the operator, $a$, 
relies on the time derivative of the thermal Bogoliubov 
parameter and written as  
\begin{eqnarray}
\Omega_p^{\alpha \beta} (t_x)
 \equiv \omega_p \delta^{\alpha \beta}
  -i \dot{n}_p(t_x) T_0^{\alpha \beta}. \label{EEVOmg}
\end{eqnarray}
Hence we write the time evolution for the positive 
frequency part of the scalar field as
\begin{eqnarray}
a_p^\alpha (t_x) = 
 {\rm exp} \left\{ -i\int_{-\infty}^{t_x} dt_s \Omega_p(t_s) 
 \right\}^{\alpha \beta}
 a_p^\beta.
 \label{AO1}
\end{eqnarray}
We obtain the differential equation
for the operator, $\bar{a}$, by the thermal Bogoliubov 
transformation (\ref{TBT2}) for Eq.~(\ref{EOMXi2}),
\begin{eqnarray}
\partial_{t_x}\bar{a}_p^\alpha (t_x) 
= \bar{a}_p^\beta (t_x)
 i (\omega_p -i \dot{n}_p(t_x)T_0)^{\beta \alpha}.
\end{eqnarray}
The time dependence for the negative frequency part 
of the scalar field is represented as
\begin{eqnarray}
 \bar{a}_p^\alpha (t_x) = \bar{a}_p^{\beta}\ 
 {\rm exp} \left\{ i \int_{-\infty}^{t_x} dt_s \Omega_p(t_s)
  \right\}^{\beta \alpha}.
 \label{AO2}
\end{eqnarray}
The positive and negative frequency parts rely on 
the same energy eigenvalue. 

Both the operators, $a_{p}$ and $\bar{a}_p$ are organized into the
positive and negative frequency parts of neutral scalar 
fields,
\begin{eqnarray}
&&  \phi_{a}^\alpha(x) \equiv \int \frac{d^3\Vec{p}}{(2\pi)^3}
 \frac{1}{\sqrt{2\omega_p}} 
\left\{ 
a_{p}^\alpha (t_x) 
e^{ i \Vec{p} \cdot \Vec{x}} 
+ (\tau_3 \bar{a}_p(t_x)^T)^\alpha 
\ e^{ -i \Vec{p} \cdot \Vec{x}} \right\}, 
\label{phia1} \\
&&  \bar{\phi}_{a}^\alpha (x) \equiv \int \frac{d^3\Vec{p}}{(2\pi)^3} 
\frac{1}{\sqrt{2\omega_p}} 
\left\{ \bar{ a}_{p}^\alpha (t_x) 
e^{ -i \Vec{p} \cdot \Vec{x}} 
+ ( a_p(t_x)^{T} \tau_3)^\alpha 
e^{ i \Vec{p} \cdot \Vec{x}} \right\}. \label{phia2}
\end{eqnarray}
Due to the non-Hermiticity of Eqs.~(\ref{AO1}) and (\ref{AO2}),
the neutral scalar fields (\ref{phia1}) and (\ref{phia2}) are not invariant
under the time-reversal transformation.
The canonical conjugate fields, $\pi_a^\alpha$ and $\bar{\pi}_a^\alpha$, are
decomposed into
\begin{eqnarray}
&&  \pi_{a}^{\alpha} (x) \equiv (-i) \int \frac{d^3\Vec{p}}{(2\pi)^3}
 \sqrt{\frac{\omega_p}{2}} 
\left\{ 
a_{p}^\alpha (t_x) \ 
e^{ i \Vec{p} \cdot \Vec{x}} 
- ( \tau_3 \bar{a}_p(t_x)^T )^\alpha 
\ e^{ -i \Vec{p} \cdot \Vec{x}} \right\}, 
\label{pia1}  \\
&&  \bar{\pi}_{a}^{\alpha}(x) \equiv (-i) \int \frac{d^3\Vec{p}}{(2\pi)^3} 
\sqrt{\frac{\omega_p}{2}} 
\left\{ -\bar{a}_{p}^\alpha (t_x)  
e^{ -i \Vec{p} \cdot \Vec{x}} 
+ ( a_p(t_x)^T \tau_3)^\alpha  
e^{ i \Vec{p} \cdot \Vec{x}} \right\}. \label{pia2}
\end{eqnarray}
Using the commutation relations (\ref{cr:a}), we calculate the equal-time 
commutation relations for $\phi_a$ and $\pi_a$ and get
\begin{eqnarray}
&& [\phi_{a}^{\alpha}(t,\Vec{x}),\pi_{a}^{\beta}(t,\Vec{y})]
 = i \delta^{(3)}(\Vec{x}-\Vec{y}) \tau_3^{\alpha \beta},\label{PhiACommRela1}\\
&& [ \bar{\phi}_{a}^{\alpha}(t,\Vec{x}),\pi_{a}^{\beta}(t,\Vec{y}) ]
 = i \delta^{(3)}(\Vec{x}-\Vec{y}) \delta^{\alpha \beta}. \label{PhiACommRela2}
\end{eqnarray}
Thus we obtain a decomposition for the scalar field by the 
operators $a$ and $\bar{a}$ with an ordinary canonical commutation 
relations.

We construct the Hamiltonian, $\hat{H}_Q$, which describes 
the time evolution of the field, $\phi_a$. 
The equations of motion for the fields, $\phi_a$ and $\pi_a$, 
are derived from Eqs.~(\ref{AO1}) and (\ref{AO2}),
\begin{eqnarray}
&& \left(1+i \frac{ \dot{n}_{|\nabla_x|}(t_x)}{\hat{\omega}_{\nabla_x}}
 T_0\right)^{\alpha \beta}
 \partial_{t_x}\phi_{a}^{\beta}(x) 
 = \pi_{a}^{\alpha}(x), \label{EOMa1} \\ 
&& \partial_{t_x}\pi_{a}^{\alpha}(x)
 = - \left(1 - i \frac{ \dot{n}_{|\nabla_x|}(t_x)}{\hat{\omega}_{\nabla_x}}T_0 
\right)^{\alpha \beta}
 (-\nabla_{x}^2 + m^2) \phi_{a}^{\beta} (x), \label{EOMa2}
\end{eqnarray}
where we make the definition, $\hat{\omega}_{\nabla_x} \equiv \sqrt{-\nabla_{x}^2+m^2}$. 
Thus the Hamiltonian, $\hat{H}_Q$, is found to be
\begin{eqnarray}
\hat{H}_Q &=& \int d^3\Vec{x} \Biggl[ 
 \frac{1}{2} \bar{\pi}_a^{\alpha}(x)
 \left( 1 - i \frac{\dot{n}_{|\nabla_x|}(t_x)}{\hat{\omega}_{\nabla_x}}
 T_0 \right)^{\alpha \beta} \pi_{a}^{\beta}(x)
 \nonumber \\ 
&& +\frac{1}{2} \bar{\phi}_a^{\alpha}(x) \left( 1 - i
 \frac{\dot{n}_{|\nabla_x|}(t_x)}{\hat{\omega}_{\nabla_x}}
 T_0 \right)^{\alpha \beta} 
 (-\nabla_x^2 + m^2)
 \phi_{a}^{\beta} (x)
\Biggr].	\label{HQ}
\end{eqnarray}
The equations of motion (\ref{EOMa1}) and (\ref{EOMa2}) are reproduced
from this Hamiltonian.

\section{Self-consistency renormalization condition}
In the previous section the Hamiltonian, $\hat{H}_Q$, is 
obtained from the equation of motion for the 
neutral scalar field, $\phi_a$. 
Here we regard the Hamiltonian, $\hat{H}_Q$, as 
the unperturbed part and adopt the perturbation theory. 
In the thermal doublet notation the quantum field theory
for a neutral scalar field is defined by the Hamiltonian, 
\begin{eqnarray}
\hat{H} = \hat{H}_0 + \hat{H}_{int},
\end{eqnarray}
where $\hat{H}_0$ and $\hat{H}_{int}$ represent 
the free and interaction part of the hat-Hamiltonian for  neutral scalar
field, respectively. 
The free hat-Hamiltonian is written as 
\begin{eqnarray}
\hat{H}_0 = 
 \int d^3\Vec{x} \Biggl[ \ \frac{1}{2} \bar{\pi}_a^{\alpha}(x) \pi_a^{\alpha} (x)
 +\frac{1}{2} \bar{\phi}_a^{\alpha}(x) (-\nabla_x^2 + m^2) \phi_a^{\alpha}(x)
 \Biggr]. \label{H0}
\end{eqnarray}
The unperturbed hat-Hamiltonian, $\hat{H}_Q$, is given from Eq.~(\ref{HQ}) by 
\begin{eqnarray}
\hat{H}_Q = \hat{H}_0 - \hat{Q}, 	\label{H}
\end{eqnarray}
where $\hat{Q}$ is called the thermal counter term and is found to be 
\begin{eqnarray}
\hat{Q} &=& \int d^3{\Vec{x}} \Biggl[ \frac{1}{2} \bar{\pi}_a^{\alpha}(x)
 i \frac{\dot{n}_{|\nabla_x|}(t_x)}{\hat{\omega}_{\nabla_x}}
 T_0^{\alpha\beta} \pi_{a}^{\beta}(x)
 \nonumber \\ 
&& + \frac{1}{2} \bar{\phi}_a^{\alpha}(x)  i
 \frac{\dot{n}_{|\nabla_x|}(t_x)}{\hat{\omega}_{\nabla_x}}
 T_0^{\alpha \beta}
 (-\nabla_x^2 + m^2)
 \phi_{a}^{\beta} (x) \Biggr].	\label{HC}
\end{eqnarray}
So the interaction hat-Hamiltonian in NETFD, denoted by $\hat{H}_I$, 
is not $\hat{H}_{int}$ but has to include the counter term, 
\begin{eqnarray}
\hat{H}_I = \hat{H}_{int} + \hat{Q},
\end{eqnarray}
and the total Hamiltonian is rewritten as
\begin{eqnarray}
\hat{H}=\hat{H}_Q + \hat{H}_I. 
\end{eqnarray}
Below we develop the perturbation theory in NETFD 
with respect to the unperturbed Hamiltonian, $\hat{H}_Q$, 
and the interaction Hamiltonian, $\hat{H}_I$. 

According to the Dyson-Wick formalism, we evaluate a quantum correlation function.
Thus the scalar propagator is given by 
\begin{eqnarray}
D_{H}^{\alpha\beta}(t_x,t_y,\Vec{x}-\Vec{y})
= \langle \theta| T[\phi_a^\alpha (x) \bar{\phi}_a^\beta (y)
u(\infty,-\infty)]|\theta\rangle,
\label{pro:1}
\end{eqnarray}
where $T$ denotes time-ordering 
and the operator $u(t,t^{\prime})$ is given by
\begin{eqnarray}
u(t, t^{\prime}) = {\exp} \left( -i\int_{t^{\prime}}^{t} dt_s \hat{H}_I (t_s) \right). 
\label{u1}
\end{eqnarray}
Below we drop the momentum label, $\Vec{p}$, in the propagator and 
the scalar field for simplicity. 
From (\ref{HI-C1}) 
the thermal bra vacuum is invariant under the time evolution,
$\langle \theta| u(t,t^\prime) = \langle \theta|$. 

The propagator (\ref{pro:1}) is rewritten in terms of the 
transformed operators, $\xi$ and $\bar{\xi}$,
\begin{eqnarray}
&& D_H^{\alpha \beta}(t_x,t_y,\Vec{x}-\Vec{y}) \nonumber \\
&&= B^{-1}(n_{|\nabla_x|} (t_x))^{\alpha \gamma_1}
\Bigl[ \theta(t_x-t_y) \langle \theta| \phi_{\xi,+}^{\gamma_1}(x)
 u(t_x,t_y) \bar{\phi}_{\xi,-}^{\gamma_2}(y)u(t_y,-\infty)|\theta\rangle  \nonumber \\
&&~~ + 
\theta (t_y-t_x) \langle \theta| \bar{\phi}_{\xi,-}^{\gamma_2}(y)
 u(t_y,t_x) \phi_{\xi,+}^{\gamma_1}(x)u(t_x,-\infty)|\theta\rangle 
\Bigl]B(n_{|\overleftarrow{\nabla}_y|} (t_y))^{\gamma_2\beta} \nonumber \\
&&+ 
B^{-1}(n_{|\nabla_x|} (t_x))^{\alpha\gamma_1}
\Bigl[  \theta (t_x-t_y) \langle \theta| \phi_{\xi,+}^{\gamma_1}(x)
  u(t_x,t_y) \{\bar{\phi}_{\xi,+}(y)\tau_3\}^{\gamma_2}
u(t_y,-\infty)|\theta\rangle	\nonumber \\
&&~~ +
\theta (t_y-t_x) \langle \theta| \{ \bar{\phi}_{\xi,+}(y) \tau_3 \}^{\gamma_2} 
 u(t_y,t_x) \phi_{\xi,+}^{\gamma_1}(x) u(t_x,-\infty) |\theta \rangle 
\Bigr]\{ B^{-1}(n_{|\overleftarrow{\nabla}_y|} (t_y))^T \tau_3 \}^{\gamma_2\beta} 
\nonumber \\
&&+ 
\{ \tau_3 B(n_{|\nabla_x|} (t_x))^T \}^{\alpha \gamma_1}\Bigl[
\theta (t_x-t_y) \langle \theta| \{ \tau_3 \phi_{\xi,-}(x) \}^{\gamma_1} 
 u(t_x,t_y) \bar{\phi}_{\xi,-}^{\gamma_2}(y) u(t_y,-\infty) |\theta \rangle \nonumber \\
&&~~ + \theta (t_y-t_x)\langle \theta| \bar{\phi}_{\xi,-}^{\gamma_2}(y)
  u(t_y,t_x) \{ \tau_3 \phi_{\xi,-}(x) \}^{\gamma_1}
u(t_x,-\infty)|\theta\rangle
 \Bigr] B(n_{|\overleftarrow{\nabla}_y|} (t_y))^{\gamma_2 \beta} \nonumber \\
&&+
  \{ \tau_3 B(n_{|\nabla_x|} (t_x))^T \}^{\alpha \gamma_1} \Bigl[ 
\theta (t_x-t_y) 
\langle \theta| \{ \tau_3 \phi_{\xi,-}(x)\}^{\gamma_1}
 u(t_x,t_y) \{ \bar{\phi}_{\xi,+}(y)\tau_3\}^{\gamma_2}
 u(t_y,-\infty) |\theta\rangle 
\nonumber \\
&&~~+ \theta (t_y-t_x)
\langle \theta| \{ \bar{\phi}_{\xi,+}(y) \tau_3 \}^{\gamma_2}
 u(t_y,t_x) \{ \tau_3 \phi_{\xi,-}(x) \}^{\gamma_1}
 u(t_x,-\infty) |\theta\rangle \Bigr]	\nonumber \\
&&~~~\times  \{ B^{-1}(n_{|\overleftarrow{\nabla}_y|} (t_y))^T \tau_3 \}^{\gamma_2\beta},
\label{pro:S}
\end{eqnarray}
where the fields $\phi_{\xi,\pm}$ and $\bar{\phi}_{\xi,\pm}$ show 
the positive and negative frequency parts, respectively,
\begin{eqnarray}
&&  \phi_{\xi,+}^{\alpha}(x) = 
 \int \frac{d^{3}\Vec{p}}{(2\pi)^3} \frac{1}{\sqrt{2\omega_p}} 
 \xi_{p}^{\alpha} 
e^{-i \omega_p t_x} e^{i \Vec{p}\cdot \Vec{x}} , \label{phixi+1} \\
&&  \phi_{\xi,-}^{\alpha}(x) = 
 \int \frac{d^{3}\Vec{p}}{(2\pi)^3} 
\frac{1}{\sqrt{2\omega_p}} 
 (\tau_3 \bar{\xi}_p^T)^\alpha 
e^{i \omega_p t_x} e^{-i \Vec{p}\cdot \Vec{x}}, \label{phixi-1} \\
&&  \bar{\phi}_{\xi,+}^{\alpha}(x) =  
\int \frac{d^{3}\Vec{p}}{(2\pi)^3} \frac{1}{\sqrt{2\omega_p}} 
 (\xi_p ^{T} \tau_3)^\alpha  e^{ -i \omega_p t_x} 
e^{i \Vec{p} \cdot \Vec{x}}, \label{phixi+2} \\
&& \bar{\phi}_{\xi,-}^{\alpha}(x) =  
\int \frac{d^{3}\Vec{p}}{(2\pi)^3} 
\frac{1}{\sqrt{2\omega_p}} \bar{\xi}_{p}^{\alpha} 
 e^{ i \omega_p t_x } e^{-i \Vec{p}\cdot \Vec{x}}. \label{phixi-2} 
\end{eqnarray}
From Eqs.~(\ref{theta1}) and (\ref{theta2}) it is found that
the propagator (\ref{pro:S}) has the following structure 
with respect to the thermal Bogoliubov matrices, 
$B(n_p(t_x))$ and $B(n_p(t_y))$, 
\begin{eqnarray}
&& D_H^{\alpha \beta} (t_{x} , t_{y}, \Vec{x}-\Vec{y}) 
 = B^{-1}(n_{|\nabla_x|} (t_x)) 
 \left(
   \begin{array}{cc}
	d_1^{11}(x,y) & d_1^{12}(x,y) \\
	0 & d_1^{22}(x,y)
   \end{array}
\right) B(n_{|\overleftarrow{\nabla}_y|}(t_y)) \nonumber \\
&& + B^{-1}(n_{|\nabla_x|} (t_x)) 
 \left(
   \begin{array}{cc}
		d_2^{11}(x,y) & d_2^{12}(x,y) \\
		d_2^{21}(x,y) & 0
   \end{array}
\right) 
 B^{-1}(n_{|\overleftarrow{\nabla}_y|}(t_y))^T \tau_3  \nonumber \\
&& + \tau_3 B (n_{|\nabla_x|} (t_x))^T  
 \left(
   \begin{array}{cc}
	0 & d_3^{12}(x,y) \\
	d_3^{21}(x,y) & d_3^{22}(x,y)
   \end{array}
\right) 
 B(n_{|\overleftarrow{\nabla}_y|}(t_y)) \nonumber \\
&& +  \tau_3 B(n_{|\nabla_x|}(t_x))^T 
 \left(
   \begin{array}{cc}
	d_4^{11} (x,y) & 0 \\
	d_4^{21} (x,y) & d_4^{22}(x,y)
   \end{array}
\right) 
   B^{-1}(n_{|\overleftarrow{\nabla}_y|} (t_y))^T \tau_3, 
\label{sp:1-2}
\end{eqnarray}
where 
\begin{eqnarray}
&& d_1^{\gamma_1 \gamma_2}(x,y) 
=  \theta(t_x-t_y) \langle \theta| \phi_{\xi,+}^{\gamma_1}(x)
 u(t_x,t_y) \bar{\phi}_{\xi,-}^{\gamma_2}(y)u(t_y,-\infty)|\theta\rangle  \nonumber \\
&&~~ + 
\theta (t_y-t_x) \langle \theta| \bar{\phi}_{\xi,-}^{\gamma_2} (y)
 u(t_y,t_x) \phi_{\xi,+}^{\gamma_1}(x) u(t_x,-\infty)|\theta\rangle, 
\label{d-1} \\
&& d_2^{\gamma_1 \gamma_2}(x,y)
=\theta (t_x-t_y) \langle \theta| \phi_{\xi,+}^{\gamma_1}(x)
  u(t_x,t_y) \{\bar{\phi}_{\xi,+}(y)\tau_3\}^{\gamma_2}
u(t_y,-\infty)|\theta\rangle	\nonumber \\
&&~~ +
\theta (t_y-t_x) \langle \theta| \{ \bar{\phi}_{\xi,+}(y) \tau_3 \}^{\gamma_2} 
 u(t_y,t_x) \phi_{\xi,+}^{\gamma_1} (x) u(t_x,-\infty) |\theta \rangle, 
\label{d-2} \\
&& d_3^{\gamma_1 \gamma_2}(x,y)
=\theta (t_x-t_y) \langle \theta| \{ \tau_3 \phi_{\xi,-}(x) \}^{\gamma_1} 
 u(t_x,t_y) \bar{\phi}_{\xi,-}^{\gamma_2}(y) u(t_y,-\infty) |\theta \rangle \nonumber \\
&&~~ + \theta (t_y-t_x)\langle \theta| \bar{\phi}_{\xi,-}^{\gamma_2}(y)
  u(t_y,t_x) \{ \tau_3 \phi_{\xi,-}(x) \}^{\gamma_1}u(t_x,-\infty)|\theta\rangle, 
\label{d-3} \\
&& d_4^{\gamma_1 \gamma_2}(x,y)
=\theta (t_x-t_y) 
\langle \theta| \{ \tau_3 \phi_{\xi,-}(x)\}^{\gamma_1}
 u(t_x,t_y) \{ \bar{\phi}_{\xi,+}(y)\tau_3\}^{\gamma_2}
 u(t_y,-\infty) |\theta\rangle 
\nonumber \\
&&~~ + \theta (t_y-t_x)
\langle \theta| \{ \bar{\phi}_{\xi,+}(y) \tau_3 \}^{\gamma_2}
 u(t_y,t_x) \{ \tau_3 \phi_{\xi,-}(x) \}^{\gamma_1}
 u(t_x,-\infty) |\theta\rangle. 
\label{d-4}
\end{eqnarray}

Performing the thermal Bogoliubov transformations for Eq.~(\ref{HC}),
we rewrite the thermal counter term $\hat{Q}$ by the transformed
operators, $\xi$ and $\bar{\xi}$,
\begin{eqnarray}
\hat{Q}= -i \int \frac{d^3\Vec{p}}{(2\pi)^3} \dot{n}_p(t_x)
\tilde{\xi}_{p}^\dagger \xi_{p}^\dagger.
\label{Q:xi}
\end{eqnarray}
The thermal counter term satisfies $\langle \theta | \hat{Q}=0$. 
This condition is important, and necessary to 
use the Feynman diagram procedure\cite{umezawa1,FynPert1}. 

The thermal counter term can be fixed by the renormalization condition. 
Since Eq.~(\ref{Q:xi}) is proportional to 
$\xi^\dagger \tilde{\xi}^\dagger$,  the thermal counter term appears in the inverse
propagator or the self energy, and modifies, at the leading order for the propagator,
only $d_1^{12}$ and $d_4^{21}$ in Eqs.~(\ref{d-1})-(\ref{d-4}). 
Substituting Eq.~(\ref{Q:xi}) into Eq.~(\ref{sp:1-2}), we obtain
\begin{eqnarray}
&& \langle \theta| T[\phi_a^\alpha (x) \bar{\phi}_a^\beta (y)
u(\infty,-\infty)]|\theta\rangle 
-\langle \theta| T[\phi_a^\alpha (x) \bar{\phi}_a^\beta (y)]|\theta\rangle  
\nonumber \\
&&= \int \frac{d^{3}\Vec{p}}{(2\pi)^3} 
\frac{1}{2\omega_p}e^{-i\omega_p \cdot (t_x-t_y)} e^{i \Vec{p}\cdot (\Vec{x}-\Vec{y})}
\nonumber \\
&&~ \times  \Biggl( \theta (t_x-t_y)\int_{-\infty}^{t_y} dt_s 
  \dot{n}_p(t_s)  
 + \theta (t_y-t_x)\int_{-\infty}^{t_x} dt_s 
 \dot{n}_p(t_s) 
 \Biggr) \nonumber \\
&&~ \times B^{-1}(n_p(t_x)) 
\left(
   \begin{array}{cc}
	0 & 1 \\
	0 & 0 \\
   \end{array}
\right) B(n_p(t_y)) \nonumber \\
&& + \int \frac{d^3\Vec{p}}{(2\pi)^3} 
\frac{1}{2\omega_p}e^{ i \omega_p \cdot (t_x-t_y)}e^{-i \Vec{p}\cdot(\Vec{x}-\Vec{y})}
\nonumber \\
&&~ \times \Biggl( \theta (t_x-t_y)\int_{-\infty}^{t_y} dt_s 
  \dot{n}_p(t_s)  
 + \theta (t_y-t_x)\int_{-\infty}^{t_x} dt_s 
  \dot{n}_p(t_s) 
 \Biggr) \nonumber \\
&&~ \times \tau_3 B(n_p(t_x))^T  
\left(
   \begin{array}{cc}
	0 & 0 \\
	1 & 0 \\
   \end{array}
\right) 
 B^{-1}(n_p(t_y))^T \tau_3. 
\label{sp:1}
\end{eqnarray}

On the other hand, the quantum correction at the leading order
is written in $2 \times 2$ matrix form by using 
the free propagator, $D_{0}^{\alpha\beta}$, 
and the self-energy, $\Sigma^{\gamma_1 \gamma_2}$, 
\begin{eqnarray}
&&\int d^4 z_1 d^4 z_2 
 D_{0}^{\alpha\gamma_1}(t_x,t_{z_1},\Vec{x}-\Vec{z_1}) 
i\Sigma^{\gamma_1 \gamma_2}(t_{z_1},t_{z_2},\Vec{z_1}-\Vec{z_2})
D_{0}^{\gamma_2 \beta}(t_{z_2},t_y,\Vec{z_2}-\Vec{y}) \nonumber \\
&& = \int d^4 z_1 d^4 z_2 B^{-1}(n_{|\nabla_x|}(t_x)) \nonumber \\
&& \ \ \times \left(
   \begin{array}{c}
	D_{0,R}^{11}(x- z_1) i\Sigma_R (t_{z_1},t_{z_2},\Vec{z_1}-\Vec{z_2}) 
D_{0,R}^{11}(z_2 - y)
 \ \ \ \ \delta \Sigma_{B,1} (x,z_1,z_2,y) \\
  \ \ \ \ \ \ \ \ \ \ \ \ 0 \ \ \ \ \ \ \ \ \ \ 
  D_{0,R}^{22}(x-z_1) i\Sigma_A (t_{z_1},t_{z_2},\Vec{z_1}-\Vec{z_2}) D_{0,R}^{22}(z_2-y) 
   \end{array}
\right) \nonumber \\
&& \ \ \ \times B(n_{|\overleftarrow{\nabla}_y|}(t_y)) \nonumber \\
&& + \int d^4 z_1 d^4 z_2  B^{-1}(n_{|\nabla_x|} (t_x)) \nonumber \\
&& \ \ \times \left(
   \begin{array}{c}
	  \delta \Sigma_{B,2} (x, z_1, z_2, y) \ \ \ \ 
	  -D_{0,R}^{11}(x - z_1)i\Sigma_R (t_{z_1},t_{z_2},\Vec{z_1}-\Vec{z_2})
D_{0,A}^{22}(z_2 - y) \\
	  -D_{0,R}^{22}(x - z_1)i\Sigma_A (t_{z_1},t_{z_2},\Vec{z_1}-\Vec{z_2})
D_{0,A}^{11}( z_2 - y)
	  \ \ \ \ \ \ \ \ \ \ 0 \ \ \ \ \ \ \ \ \ \ \ 
   \end{array}
\right) \nonumber \\
&& \ \ \ \times B^{-1}(n_{|\overleftarrow{\nabla}_y|}(t_y))^T \tau_3 \nonumber \\
&& + \int d^4 z_1 d^4 z_2 \tau_3 B(n_{|\nabla_x|} (t_x))^T \nonumber \\
&& \ \ \times \left(
   \begin{array}{c}
	  \ \ \ \ \ \ \ \ \ \ 0 \ \ \ \ \ \ \ \ 
	  -D_{0,A}^{11}(x - z_1) i\Sigma_A(t_{z_1},t_{z_2},\Vec{z_1}-\Vec{z_2})
D_{0,R}^{22}(z_2 - y) \\
	  -D_{0,A}^{22}(x - z_1) i\Sigma_R(t_{z_1},t_{z_2},\Vec{z_1}-\Vec{z_2})
D_{0,R}^{11}( z_2 - y)
	  \ \ \ \ \delta \Sigma_{B,3} ( x, z_1, z_2, y)
   \end{array}
\right) \nonumber \\
&& \ \ \ \times B(n_{|\overleftarrow{\nabla}_y|}(t_y)) \nonumber \\
&& + \int d^4 z_1 d^4 z_2  \tau_3 B(n_{|\nabla_x|} (t_x))^T \nonumber \\
&& \ \ \times \left(
   \begin{array}{c}
	D_{0,A}^{11}( x - z_1) i\Sigma_A (t_{z_1},t_{z_2},\Vec{z_1}-\Vec{z_2})
 D_{0,A}^{11} ( z_2 - y)
  \ \ \ \ \ \ \ \ 0 \ \ \ \ \ \ \ \ \ \ \  \\
  \delta \Sigma_{B,4} (x,z_1,z_2,y)  \ \ \ \ 
  D_{0,A}^{22}( x - z_1) i\Sigma_R (t_{z_1},t_{z_2},\Vec{z_1}-\Vec{z_2})
 D_{0,A}^{22}(z_2 - y)    
   \end{array}
\right) \nonumber \\
&& \ \ \ \times B^{-1}(n_{|\overleftarrow{\nabla}_y|}(t_y))^T \tau_3, 
\label{sp:2}
\end{eqnarray}
with
\begin{eqnarray}
&& \delta \Sigma_{B,1}( x, z_1, z_2,y)  \nonumber \\
&& = D_{0,R}^{11}(x - z_1)
 \Bigl\{ i\Sigma^{12} (t_{z_1},t_{z_2},\Vec{z_1}-\Vec{z_2}) 
+ i \Sigma_R (t_{z_1},t_{z_2},\Vec{z_1}-\Vec{z_2}) 
n_{|\overleftarrow{\nabla}_{z_2}|} (t_{z_2}) \nonumber \\
&& - n_{|\nabla_{z_1}|} (t_{z_1}) i \Sigma_A (t_{z_1},t_{z_2},\Vec{z_1}-\Vec{z_2}) \Bigr\}
 D_{0,R}^{22}(z_2 - y),  \label{SEND1} \\
&& \delta \Sigma_{B,2} ( x, z_1, z_2, y) \nonumber \\
&&= D_{0,R}^{11}(x - z_1)
 \Bigl\{ i\Sigma^{11} (t_{z_1},t_{z_2},\Vec{z_1}-\Vec{z_2}) 
 + i \Sigma_R(t_{z_1}-t_{z_2},\Vec{z_1}-\Vec{z_2}) 
n_{|\overleftarrow{\nabla}_{z_2}|} (t_{z_2}) \nonumber \\
&& + n_{|\nabla_{z_1}|} (t_{z_1}) i \Sigma_A(t_{z_1},t_{z_2},\Vec{z_1}-\Vec{z_2}) \Bigr\}
 D_{0,A}^{11}( z_2 -  y),  \label{SEND2} \\
&& \delta \Sigma_{B,3}( x, z_1, z_2, y) \nonumber \\
&&= -D_{0,A}^{22}( x - z_1) 
 \Bigl\{ i\Sigma^{22} (t_{z_1},t_{z_2},\Vec{z_1}-\Vec{z_2}) 
 + i \Sigma_R (t_{z_1},t_{z_2},\Vec{z_1}-\Vec{z_2}) 
 n_{|\overleftarrow{\nabla}_{z_2}|} (t_{z_2}) \nonumber \\
&& + n_{|\nabla_{z_1}|} (t_{z_1}) i \Sigma_A (t_{z_1},t_{z_2},\Vec{z_1}-\Vec{z_2}) \Bigr\}
 D_{0,R}^{22}( z_2 - y),  \label{SEND3} \\
&& \delta \Sigma_{B,4} ( x, z_1, z_2, y) \nonumber \\
&&= D_{0,A}^{22}( x - z_1)
 \Bigl\{ -i\Sigma^{21} (t_{z_1},t_{z_2},\Vec{z_1}-\Vec{z_2}) 
 - i \Sigma_R(t_{z_1},t_{z_2},\Vec{z_1}-\Vec{z_2}) 
 n_{|\overleftarrow{\nabla}_{z_2}|} (t_{z_2}) \nonumber \\
&& + n_{|\nabla_{z_1}|} (t_{z_1}) i \Sigma_A(t_{z_1},t_{z_2},\Vec{z_1}-\Vec{z_2}) \Bigr\}
 D_{0,A}^{11}( z_2 - y). \label{SEND4}
\end{eqnarray}
An explicit form for the free propagator is given in (\ref{ThermPro0}).
The retarded and advanced parts of the self-energy, 
$\Sigma_R$ and $\Sigma_A$, are defined by
\begin{eqnarray}
\Sigma_R \equiv \Sigma^{11}+\Sigma^{12} = \Sigma^{21}+\Sigma^{22},
 \ \ \ \ 
\Sigma_A \equiv  \Sigma^{11}-\Sigma^{21} = \Sigma^{22}-\Sigma^{12}.
\label{SelfERA}
\end{eqnarray}
The first and  last terms in the right-hand side of 
Eq.~(\ref{sp:2}) have the same Bogoliubov transformation
structure as the first and last ones in the 
right-hand side of Eq.~(\ref{sp:1}), respectively. 
Below we identify Eqs.~(\ref{sp:1}) and (\ref{sp:2})
as the thermal counter terms and the contribution of
quantum corrections, respectively.

H.~Chu and H.~Umezawa have proposed the self-consistency 
renormalization condition to fix the thermal counter term. 
The condition imposes 
$\langle \theta|\xi_{H,p} (t_x)\tilde{\xi}_{H,p}(t_x)|\theta\rangle=0$ 
at the equal time limit, $t_x \rightarrow t_y$ \cite{SCRC2}, 
where the subscript $H$ denotes the Heisenberg picture and whose implication
will be seen at the top of the next section. 
The condition amounts to the vanishing off-diagonal elements, $d_1^{12}$ and $d_4^{21}$,
in the limit. 
\footnote{
Substituting the fields (\ref{phixi+1})-(\ref{phixi-2}) and taking the equal time limit, 
we obtain
\begin{eqnarray*}
\lim_{t_x\rightarrow t_y}
d_1^{12}(x,y)&=& -\int \frac{d^3 \Vec{p}}{(2\pi)^3} \frac{d^3 \Vec{k}}{(2\pi)^3}
\frac{1}{\sqrt[]{2\omega_p}} \frac{1}{\sqrt[]{2\omega_k}} 
e^{ i \Vec{p} \cdot \Vec{x}} e^{ -i \Vec{k} \cdot \Vec{y}} 
\langle \theta|  T[\xi_p(t_x) \tilde{\xi}_k(t_x)u(\infty, -\infty)] |\theta\rangle,\\
\lim_{t_x\rightarrow t_y}
d_4^{21}(x,y)
&=& -\int \frac{d^3 \Vec{p}}{(2\pi)^3} \frac{d^3 \Vec{k}}{(2\pi)^3}
\frac{1}{\sqrt[]{2\omega_p}} \frac{1}{\sqrt[]{2\omega_k}} 
e^{ -i \Vec{p} \cdot \Vec{x}} e^{ i \Vec{k} \cdot \Vec{y}} 
\langle \theta| T[ \tilde\xi_p (t_x) \xi_k (t_x)u(\infty, -\infty)] |\theta\rangle.
\end{eqnarray*}
}
Due to the tilde conjugation rules, both the equations
give an equivalent condition. Thus the self-consistency 
renormalization conditions reduce to a single equation.
From Eqs.~(\ref{sp:1}) and Eq.~(\ref{sp:2}) we obtain
\begin{eqnarray}
&&\int_{-\infty}^{t_x} dt_s \int \frac{d^3 \Vec{p}}{(2\pi)^3} 
\frac{1}{2\omega_p} \dot{n}_p(t_s) 
e^{i\Vec{p}\cdot (\Vec{x}-\Vec{y}) } \nonumber \\
&&~~ +\lim_{t_x\rightarrow t_y}\int d^4 z_1 d^4 z_2
 \delta \Sigma_{B,1}( x, z_1, z_2, y)=0. \label{Self-C1} 
\end{eqnarray}
In thermal equilibrium
$\delta \Sigma_{B,1}$ and $\delta \Sigma_{B,4}$ vanish, so that
these conditions are satisfied automatically. 
In NETFD Eq.~(\ref{Self-C1}) shows 
the time evolution for the thermal Bogoliubov parameter, ${n}_p(t)$.

For a practical calculation in homogeneous NETFD 
it is more convenient to employ the $t$-representation, while  
the spatial Fourier transformation is performed. 
Differentiating Eq.~(\ref{Self-C1}) with respect to $t_x$ 
and performing the spatial Fourier transformation, we obtain
\begin{eqnarray}
\dot{n}_p(t_x) =
-2 \omega_p \partial_{t_x} \Bigl\{ \lim_{t_x\rightarrow t_y}\int dt_{z_1}dt_{z_2}
  \delta \Sigma_{B,1}(t_x,t_{z_1},t_{z_2},t_y;\Vec{p}) \Bigr\}, \label{BoltzEq1} 
\end{eqnarray}
where $\delta \Sigma_{B,1}(t_x,t_{z_1},t_{z_2},t_y;\Vec{p})$ is 
written by the propagator and the self-energy in the $t$-representation, 
\begin{eqnarray}
&& \delta \Sigma_{B,1}(t_x,t_{z_1},t_{z_2},t_y;\Vec{p})  
= D_{0,R}^{11}(t_x - t_{z_1};\Vec{p})
 \Bigl\{ i\Sigma^{12} (t_{z_1},t_{z_2},\Vec{p}) \nonumber \\
&&~~ + n_p(t_{z_2}) i \Sigma_R (t_{z_1},t_{z_2},\Vec{p}) 
 - n_p(t_{z_1}) i \Sigma_A (t_{z_1},t_{z_2},\Vec{p}) \Bigr\}
 D_{0,R}^{22}(t_{z_2} - t_y;\Vec{p}). \label{ns_t-rep_SEND1}
\end{eqnarray}
This equation corresponds to the quantum Boltzmann equation for a
relativistic neutral scalar field. 

\section{Boltzmann equation for a neutral scalar field}

The Heisenberg number density, $n_{H,p}(t)$, is defined by
\begin{eqnarray}
 (2\pi)^3\delta^{(3)}(\Vec{p}-\Vec{k}) n_{H,p}(t)
=\langle \theta|a_{H,p}^\dagger (t) a_{H,k}(t) |\theta \rangle.
\end{eqnarray}
It is the self-consistency condition\cite{umezawa1,pertN1,pertN2} 
in the previous section that 
establishes the correspondence between the thermal Bogoliubov parameter, $n_p(t)$, 
and the above particle number density, $n_{H,p}(t)$.
We can confirm the correspondence as follows.
The interaction hat-Hamiltonian, $\hat{H}_I$, annihilates the bra vacuum 
(but not the ket vacuum in general though),
\begin{eqnarray}
	\langle \theta| \hat{H}_I = 0,\qquad 
	\hat{H}_I |\theta \rangle \ne 0 ,
\end{eqnarray}
according to the thermal state conditions Eqs.~(\ref{ns_tsc_1})-(\ref{ns_tsc_4}) 
and the specific form of the thermal counter term Eq.~(\ref{Q:xi}). 
The $\xi$-operators in the Heisenberg picture are defined by
\begin{eqnarray}
&&\xi_{H,p}^\alpha (t) 
= B(n_p(t))^{\alpha \beta} a_{H,p}^{\beta} (t), \label{HTBT1} \\
&&\bar{\xi}_{H,p}^\alpha (t)
= \bar{a}_{H,p}^{\beta} (t) B^{-1}(n_p(t))^{\beta \alpha}, \label{HBT2}
\end{eqnarray}
with the Bogoliubov parameter $n_p(t)$ (not $n_{H,p}(t)$) and  satisfy
\begin{eqnarray}
	\langle \theta| \xi_{H,p}^\dagger(t) = \langle \theta| \tilde\xi_{H,p}^\dagger(t) = 0, \qquad 
	\xi_{H,p}(t) |\theta \rangle \ne 0 ,\qquad
	\tilde\xi_{H,p}(t) |\theta \rangle \ne 0 .
\end{eqnarray}
Then it follows that
\begin{eqnarray}
	&&\langle \theta| a_{H,p}^\dagger(t) a_{H,k}(t) |\theta \rangle \nonumber\\
	&=& \langle \theta|\left\{\tilde\xi_{H,p}(t) + \left(1+n_p(t)\right)\xi^\dagger_{H,p}(t) \right\} 
	\left\{\xi_{H,k}(t) + n_k(t)\tilde\xi_{H,k}^\dagger(t) \right\} |\theta \rangle \nonumber\\
	&=& \langle \theta| \tilde\xi_{H,p}(t) \xi_{H,k}(t) |\theta \rangle + 
	n_k(t) \langle \theta| \tilde\xi_{H,p}(t) \tilde\xi_{H,k}^\dagger(t) |\theta \rangle,
\end{eqnarray}
which implies
\begin{eqnarray}
	(2\pi)^3\delta^{(3)}(\Vec{p}-\Vec{k})\bigl(n_{H,p}(t) - n_p(t)\bigr) =  
	\langle \theta| \tilde\xi_{H,p}(t) \xi_{H,k}(t) |\theta \rangle. \label{ex_1}
\end{eqnarray}
Thus the thermal Bogoliubov parameter is equal to the Heisenberg number distribution 
at each instant of time under the self-consistency renormalization condition, 
$\langle \theta|\xi_{H,p}(t) \tilde{\xi}_{H,p}(t)|\theta\rangle=0$.
In what follows we apply the self-consistency renormalization 
condition to the self-interacting systems of relativistic scalar field and 
derive the quantum Boltzmann equation for the thermal Bogoliubov parameter. 

\subsection{$\lambda\phi^3$ interaction model}
First we calculate the time evolution of the thermal
Bogoliubov parameter for a neutral scalar field with
a three-point self-interaction. We start from
the Hamiltonian,
\begin{eqnarray}
\hat{H} = H-\tilde{H}, \label{mdl1}
\end{eqnarray}
with
\begin{eqnarray}
H= \int d^3\Vec{x} \Bigl[ \frac{1}{2} \bigl\{  \pi_a (x)^2
 + \phi_a(x) (-\nabla_x^2 + m^2) \phi_a(x) \bigr\}
 + \frac{\lambda}{3!}\phi_a(x)^3\Bigr], \label{mdl-P3}
\end{eqnarray}
where $\phi_a$ and $\pi_a$ mean 
$\phi_a^1$ in Eq.~(\ref{phia1}) 
and $\pi_a^1$ in Eq.~(\ref{pia1}), 
respectively. The fields,
$\phi_a$ and $\pi_a$, are also equivalent 
to $\bar{\phi}_a^1$ in Eq.~(\ref{phia2}) 
and $\bar{\pi}_a^1$ in Eq.~(\ref{pia2}). 
The tilde conjugate Hamiltonian, $\tilde{H}$,
is described by the fields, 
$\tilde{\phi}_a$ and $\tilde{\pi}_a$,
which are equivalent to $\phi_a^2$ and 
$\pi_a^2$, respectively.
We evaluate the one-loop thermal self-energy by using 
the Feynman rules in the thermal doublet 
notation\cite{FynRule1,prop1}. 
In the $t$-representation we assign the propagator 
(\ref{ThermPro0}) to each internal line and
\begin{eqnarray}
\lambda^{\alpha} = \lambda
\left(
   \begin{array}{c}
	1 \\
	-1 \\
   \end{array}
\right ),
\label{TCC0}
\end{eqnarray}
to each vertex.

\begin{figure}[bt]
	\begin{center}
	\includegraphics{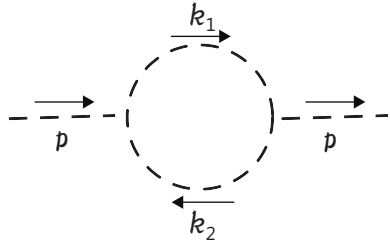}
	\end{center}
	\caption{1-loop thermal self-energy derived three point scalar interaction model}
	\label{figTSE1}
\end{figure}
At the one-loop level the thermal self-energy is 
diagrammatically represented in Fig.~\ref{figTSE1}, and is calculated as 
\begin{eqnarray}
&&i\Sigma_{B,1-loop}^{\gamma_1 \gamma_2}(t_{z_1},t_{z_2};\Vec{p}) \nonumber \\
&&= -\frac{1}{2}\int \frac{d^3 \Vec{k}_1}{(2\pi)^3} \frac{d^3 \Vec{k}_2}{(2\pi)^3} 
(2\pi)^3 \delta^{(3)} (\Vec{p} - \Vec{k}_1 + \Vec{k}_2) \nonumber \\
&&~~ \times	\tau_3^{\gamma_1 \delta} \lambda^{\delta} 
\{ D_{0}(t_{z_1},t_{z_2};\Vec{k}_1) \tau_3 \}^{\delta \gamma_2} 
\lambda^{\gamma_2}
\{ D_{0}(t_{z_2},t_{z_1};\Vec{k}_2) \tau_3 \}^{\gamma_2 \delta}  \nonumber \\
&&=-\frac{\lambda^2}{2} \sum^{2}_{i_1 =1}\sum^{2}_{i_2 =1}
\int \frac{d^3 \Vec{k}_1}{(2\pi)^3} \frac{d^3 \Vec{k}_2}{(2\pi)^3}
\frac{1}{4\omega_{k_1} \omega_{k_2}} 
(2\pi)^3 \delta^{(3)}(\Vec{p} - \Vec{k}_1 + \Vec{k}_2)\nonumber \\
&&\times \Biggl[ \theta( t_{z_1}-t_{z_2} ) \  
e^{ i(E_{k_1,i_1}+E_{k_2,i_2}) (t_{z_1}-t_{z_2})} \nonumber \\
&&\ \ \ \ \  \times \left(
   \begin{array}{cc}
	f_{k_1,i_1}(t_{z_2}) f_{k_2,i_2}(t_{z_2}) 
 & -\bar{f}_{k_1,i_1}(t_{z_2}) \bar{f}_{k_2,i_2}(t_{z_2})  \\
	f_{k_1,i_1}(t_{z_2}) f_{k_2,i_2}(t_{z_2}) 
 & -\bar{f}_{k_1,i_1}(t_{z_2}) \bar{f}_{k_2,i_2}(t_{z_2})  \\
   \end{array}
\right) \nonumber \\
&& \ \ + \theta( t_{z_2}-t_{z_1} ) \ 
e^{-i(E_{k_1,i_1} + E_{k_2,i_2}) (t_{z_1}-t_{z_2})} \nonumber \\
&&\ \ \ \ \  \times \left(
   \begin{array}{cc}
	f_{k_1,i_1}(t_{z_1}) f_{k_2,i_2}(t_{z_1}) 
 & -f_{k_1,i_1}(t_{z_1}) f_{k_2,i_2}(t_{z_1})  \\
	\bar{f}_{k_1,i_1}(t_{z_1}) \bar{f}_{k_2,i_2}(t_{z_1}) 
 & -\bar{f}_{k_1,i_1}(t_{z_1}) \bar{f}_{k_2,i_2}(t_{z_1})  \\
   \end{array}
\right) 
\Biggr],
\label{tse:1}
\end{eqnarray}
where
\begin{eqnarray}
&& E_{q,1}=\omega_q,\ \ E_{q,2} = -\omega_q, \\
&& f_{q,1}(t) = n_{q}(t),\ \ \ f_{q,2}(t) = 1+n_{q}(t), \\
&& \bar{f}_{q,1}(t) = 1+n_{q}(t),\ \ \ \bar{f}_{q,2}(t) = n_{q}(t).
\end{eqnarray}
Thus the off-diagonal elements, (\ref{SEND1}) and (\ref{SEND4}), are given by
\begin{eqnarray}
&& \lim_{t_x\rightarrow t_y}\int dt_{z_1}dt_{z_2} 
\delta \Sigma_{B,1}(t_x,t_{z_1},t_{z_2},t_y;\Vec{p}) \nonumber \\
&& = \lim_{t_x\rightarrow t_y}\int dt_{z_1}dt_{z_2} 
\delta \Sigma_{B,4}(t_x,t_{z_1},t_{z_2},t_y;\Vec{p}) \nonumber \\
&& = \int_{-\infty}^{t_x} dt_s \frac{1}{2\omega_p}
\sum^{2}_{i_1 =1}\sum^{2}_{i_2 =1} \frac{\lambda^2}{2}
\int \frac{d^3\Vec{k}_1}{(2\pi)^3} \frac{d^3\Vec{k}_2}{(2\pi)^3}   
\frac{1}{4\omega_p \omega_{k_1} \omega_{k_2}} 
(2\pi)^3 \delta^{(3)}(\Vec{p} - \Vec{k}_1 + \Vec{k}_2) \nonumber \\
&&\ \ \times \frac{{\rm sin}\{ (\omega_{p}+E_{k_1,i_1}+E_{k_2,i_2} )
 (t_x-t_s) \} }{\omega_{p}+E_{k_1,i_1}+E_{k_2,i_2}} \nonumber \\
 &&\  \times \Bigl\{ n_{p}(t_s) f_{k_1,i_1}(t_s) f_{k_2,i_2}(t_s)
 - (1+n_{p}(t_s)) \bar{f}_{k_1,i_1}(t_s) \bar{f}_{k_2,i_2}(t_s)  \Bigr\},
\label{off-dia}
\end{eqnarray}
at the equal time limit. 
When the Bose distribution function is assumed for $n_p(t)$,
Eq.~(\ref{off-dia}) vanishes. 
It shows that the Bose distribution is a stationary solution 
for Eq.~(\ref{BoltzEq1}). 

Inserting Eq.~(\ref{off-dia}) into Eq.~(\ref{BoltzEq1}), 
we obtain the time evolution equation for 
the thermal Bogoliubov parameter,
\begin{eqnarray}
&&\dot{n}_p(t_x)
 = -\frac{\lambda^2}{2} \sum^{2}_{i_1 =1}\sum^{2}_{i_2 =1} 
\int_{-\infty}^{t_x} dt_s
\int \frac{d^3 \Vec{k}_1}{(2\pi)^3} \frac{d^3 \Vec{k}_2}{(2\pi)^3} 
\frac{1}{4\omega_p \omega_{k_1} \omega_{k_2}} \nonumber \\
&&~ \times {\cos} \{ (\omega_{p}+E_{k_1,i_1}+E_{k_2,i_2} ) (t_x-t_s) \} 
(2\pi)^3 \delta^{(3)}(\Vec{p} - \Vec{k}_1 + \Vec{k}_2) \nonumber \\
 &&~ \times \Bigl\{ n_{p}(t_s) f_{k_1,i_1}(t_s) f_{k_2,i_2}(t_s)
 - (1+n_{p}(t_s)) \bar{f}_{k_1,i_1}(t_s) \bar{f}_{k_2,i_2}(t_s)  \Bigr\}.
\label{coll:bol}
\end{eqnarray}
The third line of this equation has the same
statistical structure as the quantum Boltzmann equation. 
Off-shell contribution is included through the second line of Eq.~(\ref{coll:bol}). 

At the limit, $t_x \rightarrow \infty$, the second line of
Eq.~(\ref{off-dia}) reduces to the delta function which 
guarantees the energy conservation.
Then the collision term of the quantum Boltzmann equation
is derived from Eq.~(\ref{off-dia}).
It shows that time evolution for the thermal Bogoliubov 
parameter is described by the quantum Boltzmann equation
at the limit.

\subsection{$\lambda\phi^4$ interaction model}
Next we consider a neutral scalar field with
a four-point self-interaction. The model is defined 
by the Hamiltonian,
\begin{eqnarray}
\hat{H} = H-\tilde{H},
\end{eqnarray}
with
\begin{eqnarray}
H= \int d^3\Vec{x} \Bigl[\frac{1}{2} \bigl\{  \pi_a (x)^2
 + \phi_a(x) (-\nabla_x^2 + m^2) \phi_a(x) \bigr\}
 + \frac{\lambda}{4!}\phi_a(x)^4 \Bigr]. 	\label{mdl-P4}
\end{eqnarray}
The numerical factor in the interaction term provides the 
same assignment to each vertex as in the previous model.

We calculate the thermal self-energy in this model.
\begin{figure}[b]
	\begin{center}
	\includegraphics{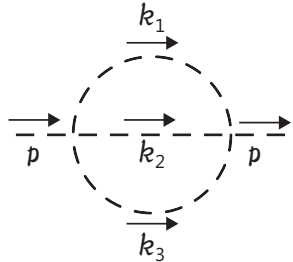}
	\end{center}
	\caption{2-loop thermal self-energy in $\lambda \phi^4$ interaction model}
	\label{figTSE2}
\end{figure}
Since there is no momentum transfer from the external
to the internal lines, the self-energy has a diagonal 
form at the one-loop level. Thus the self-consistency
renormalization condition is satisfied for 
$\dot{n}_p(t)=0$ at one-loop level.
The time evolution of the thermal Bogoliubov parameter
is induced from the two-loop self-energy illustrated in 
Fig.~\ref{figTSE2}. We compute the diagram and obtain
\begin{eqnarray}
&&i\Sigma_{B,2-loop}^{\gamma_1\gamma_2}(t_{z_1},t_{z_2};\Vec{p}) \nonumber \\
&&= -\frac{1}{3!}\int \frac{d^3 \Vec{k}_1}{(2\pi)^3}
	\frac{d^3 \Vec{k}_2}{(2\pi)^3} \frac{d^3 \Vec{k}_3}{(2\pi)^3}
	(2\pi)^3 \delta^{(3)}(\Vec{p} - \Vec{k}_1 - \Vec{k}_2 - \Vec{k}_3 ) 
\nonumber \\
&& \times \tau_3^{\gamma_1 \delta} \lambda^{\delta} 
\{ D_{0} (t_{z_1},t_{z_2};\Vec{k}_1) \tau_3 \}^{\delta \gamma_2}
\{ D_{0} (t_{z_1},t_{z_2};\Vec{k}_2) \tau_3 \}^{\delta \gamma_2}
\{ D_{0} (t_{z_1},t_{z_2};\Vec{k}_3) \tau_3\}^{\delta \gamma_2} \lambda^{\gamma_2}
\nonumber \\
&&=-\frac{\lambda^2}{3!} \sum^{2}_{i_1 =1}\sum^{2}_{i_2 =1}\sum^{2}_{i_3 =1}
\int \frac{d^3\Vec{k}_1}{(2\pi)^3} \frac{d^3\Vec{k}_2}{(2\pi)^3} 
\frac{d^3\Vec{k}_3}{(2\pi)^3} 
\frac{1}{8\omega_{k_1}\omega_{k_2}\omega_{k_3}} 
	(2\pi)^3 \delta^{(3)}(\Vec{p} - \Vec{k}_1 - \Vec{k}_2 - \Vec{k}_3 ) 
\nonumber \\
\ \ \ &&\times \Biggl[ \theta( t_{z_1}-t_{z_2} ) 
e^{i(E_{k_1,i_1}+E_{k_2,i_2}+E_{k_3,i_3})(t_{z_1}-t_{z_2})} \nonumber \\
&& \times \left(
   \begin{array}{cc}
	f_{k_1,i_1}(t_{z_2}) f_{k_2,i_2}(t_{z_2}) f_{k_3,i_3}(t_{z_2})
 & -\bar{f}_{k_1,i_1}(t_{z_2}) \bar{f}_{k_2,i_2}(t_{z_2}) \bar{f}_{k_3,i_3}(t_{z_2}) \\
	f_{k_1,i_1}(t_{z_2}) f_{k_2,i_2}(t_{z_2}) f_{k_3,i_3}(t_{z_2})
 & -\bar{f}_{k_1,i_1}(t_{z_2}) \bar{f}_{k_2,i_2}(t_{z_2}) \bar{f}_{k_3,i_3}(t_{z_2}) \\
   \end{array}
\right) \nonumber \\
&& \ \ \ + \theta( t_{z_2}-t_{z_1} ) 
e^{-i(E_{k_1,i_1}+E_{k_2,i_2}+E_{k_3,i_3})(t_{z_1}-t_{z_2})} \nonumber \\
&& \times \left(
   \begin{array}{cc}
	f_{k_1,i_1}(t_{z_1}) f_{k_2,i_2}(t_{z_1}) f_{k_3,i_3}(t_{z_1})
 & -f_{k_1,i_1}(t_{z_1}) f_{k_2,i_2}(t_{z_1}) f_{k_3,i_3}(t_{z_1}) \\
	\bar{f}_{k_1,i_1}(t_{z_1}) \bar{f}_{k_2,i_2}(t_{z_1}) \bar{f}_{k_3,i_3}(t_{z_1})
 & -\bar{f}_{k_1,i_1}(t_{z_1}) \bar{f}_{k_2,i_2}(t_{z_1}) \bar{f}_{k_3,i_3}(t_{z_1}) \\
   \end{array}
\right) 
\Biggr]. 
\label{BoseSelfE2}
\end{eqnarray}
The off-diagonal elements, (\ref{SEND1}) and 
(\ref{SEND4}), are given by
\begin{eqnarray}
&& \lim_{t_x\rightarrow t_y}\int dt_{z_1}dt_{z_2} 
\delta \Sigma_{B,1}(t_x,t_{z_1},t_{z_2},t_y;\Vec{p}) \nonumber \\
&& = \lim_{t_x\rightarrow t_y}\int dt_{z_1}dt_{z_2} 
\delta \Sigma_{B,4}(t_x,t_{z_1},t_{z_2},t_y;\Vec{p}) \nonumber \\
&&= \int_{-\infty}^{t_x}dt_s \frac{1}{2\omega_p}
 \frac{\lambda^2}{3!} \sum^{2}_{i_1 =1}\sum^{2}_{i_2 =1}\sum^{2}_{i_3 =1}
 \int \frac{d^3\Vec{k}_1}{(2{\pi})^3} \frac{d^3\Vec{k}_2}{(2{\pi})^3} 
\frac{d^3\Vec{k}_3}{(2{\pi})^3}
	\frac{1}{8\omega_{p} \omega_{k_1} \omega_{k_2} \omega_{k_3} }
 \nonumber \\
&&\times \frac{{\rm sin} \bigl\{ (\omega_{p}+E_{k_1,i_1}+E_{k_2,i_2}+E_{k_3,i_3}) (t_x-t_s)
  \bigr\}}
 { \omega_{p}+E_{k_1,i_1}+E_{k_2,i_2}+E_{k_3,i_3} }	
 (2\pi)^3 \delta^{(3)}(\Vec{p} - \Vec{k}_1 - \Vec{k}_2 - \Vec{k}_3 ) \nonumber \\
&&\times \Bigl[ n_{p}(t_s)
	 f_{k_1,i_1}(t_s) f_{k_2,i_2}(t_s) f_{k_3,i_3}(t_s)
  - (1+n_p(t_s)) \bar{f}_{k_1,i_1}(t_s) \bar{f}_{k_2,i_2}(t_s) \bar{f}_{k_3,i_3}(t_s) \Bigr], \nonumber \\
	\label{BSDOffShlKai2}
\end{eqnarray}
at the equal time limit. Substituting Eq.~(\ref{BSDOffShlKai2}) to 
Eq.~(\ref{BoltzEq1}), we obtain the time 
evolution equation for the thermal Bogoliubov parameter, 
\begin{eqnarray}
&&\dot{n}_{p}(t_x) = 
 (-1)\frac{\lambda^2}{3!} \sum^{2}_{i_1 =1}\sum^{2}_{i_2 =1}\sum^{2}_{i_3 =1}
 \int_{-\infty}^{t_x} dt_s \int \frac{d^3\Vec{k}_1}{(2{\pi})^3} \frac{d^3\Vec{k}_2}{(2{\pi})^3}
 \frac{d^3\Vec{k}_3}{(2{\pi})^3} 
	\frac{1}{8\omega_{p} \omega_{k_1} \omega_{k_2} \omega_{k_3} }	\nonumber \\
&&\times {\cos} \bigl\{ (\omega_{p}+E_{k_1,i_1}+E_{k_2,i_2}+E_{k_3,i_3}) (t_x-t_s) \bigr\}	
 (2\pi)^3 \delta^{(3)}(\Vec{p} - \Vec{k}_1 - \Vec{k}_2 - \Vec{k}_3 )
 \nonumber \\
&&\times \Bigl[ n_{p}(t_s)
	 f_{k_1,i_1}(t_s) f_{k_2,i_2}(t_s) f_{k_3,i_3}(t_s)
  - (1+n_p(t_s)) \bar{f}_{k_1,i_1}(t_s) \bar{f}_{k_2,i_2}(t_s) \bar{f}_{k_3,i_3}(t_s) \Bigr]. 
\nonumber \\	\label{BoltzEq2-loop}
\end{eqnarray}
This equation has the same statistical structure as 
the quantum Boltzmann equation for the $\lambda\phi^4$ interaction model.
It should be noticed that a coefficient of the right-hand 
side in Eq.~(\ref{BoltzEq2-loop}) is twice of the one obtained 
in Ref.~\citen{SD-2}. 
As is shown in Appendix B, Eq.~(\ref{BoltzEq2-loop}) coincides
with the quantum transport equation in the non-relativistic 
regime\cite{ColdAtm3}.

\section{Conclusion}
We have investigated a relativistic neutral
scalar field in NETFD. Thermal degree of freedom
is introduced through the time dependent 
Bogoliubov transformation. 
Then the thermal counter term has to be introduced 
for a consistent description of both $a(t)$ and $\xi(t)$
under the transformation, and is a part of the interaction 
Hamiltonian.  We adopt the perturbative expansion, and 
calculate the full propagator in the canonical 
formalism. The Bogoliubov matrix structure of the propagator is crucial. 

Applying the self-consistency renormalization 
condition\cite{SCRC2} to a relativistic neutral scalar field, 
we have derived the time evolution equation for the thermal 
Bogoliubov parameter which is considered
to be the particle number density. 
It has been shown that the equation reduces to the 
quantum Boltzmann equation in $\phi^3$ and $\phi^4$ 
interaction models.

In this paper we impose the Lorentz 
covariance for the neutral scalar field and 
decompose it in terms of the creation and 
annihilation operators, $\xi$ and $\xi^\dagger$. 
It is not always possible to do so in a general situation of
non-equilibrium system. Some modification would be necessary to apply 
the procedure to a field with a time-dependent 
screening mass, for example. 

There are some remaining problems. 
There is no counter term for the second and third lines in Eq.~(\ref{sp:1-2}). 
These terms may have a nontrivial contribution 
to the time evolution equation at higher order.@
 In the present paper we have assumed spatial homogeneity. 
The space-time dependence is also important 
to study some relativistic systems. 
Some works to extend NETFD to spatially inhomogeneous systems  have been attempted
for non-relativistic field. An essence of such extension is to expand 
the field operator not by a complete set of plane wave functions, but by 
a complete set mixing momentum for diffusion process\cite{inhomo1,inhomo2}
or by a complete set of wave functions under trapping potential for 
cold atomic system\cite{ColdAtm3,ColdAtm4}, while 
the equal-time commutation relations are preserved.
We can formulate inhomogeneous TFD for relativistic fields in similar ways. 
We are also interested in applying the procedure 
to a relativistic Dirac field and an inhomogeneous 
system. We hope to solve these problems and report
the result in future.

\section*{Acknowledgements}
Discussions during the YITP workshop 
on "Thermal Quantum Field Theories 
and Their Applications 2010" were useful to complete this work. 

\appendix

\section{Propagator for a free neutral scalar field}
In TFD the Feynman propagator for a free neutral scalar 
field is given by the expectation value of the time
ordered product of two scalar fields. It has the $2\times 2$ matrix
form in the thermal doublet notation,\cite{umezawa1}
\begin{eqnarray}
D_0^{\alpha\beta}(t_x,t_y,\Vec{x}-\Vec{y})\equiv\langle \theta|
  T[\phi_a^{\alpha}(x)\bar{\phi}_a^{\beta}(y)]|\theta\rangle,
\label{pro:a1}
\end{eqnarray}
where the neutral scalar fields, $\phi_a$ and $\bar{\phi}_a$, 
are decomposed into the positive and negative frequency
parts by Eqs.~(\ref{phia1}) and (\ref{phia2}).
The thermal Bogoliubov transformation is applied, then the propagator
(\ref{pro:a1}) reads
\begin{eqnarray}
&&D_0^{\alpha\beta}(t_x,t_y,\Vec{x}-\Vec{y}) \nonumber \\
&&= \theta(t_x-t_y)
\Bigl[ B^{-1}(n_{| \nabla_x| }(t_x))^{\alpha\gamma_1}\langle \theta|
 \phi_{\xi,+}^{\gamma_1} (x)\bar{\phi}_{\xi,-}^{\gamma_2}(y)|\theta\rangle
 B(n_{ |\overleftarrow{\nabla}_y| } (t_y))^{\gamma_2 \beta} \nonumber \\
&&~~+\{\tau_3 B(n_{|\nabla_x|}(t_x))^{T}\}^{\alpha\gamma_1}
\langle \theta| \phi_{\xi,-}^{\gamma_1}(x)\bar{\phi}_{\xi,+}^{\gamma_2}(y)
|\theta\rangle \{\tau_3 B^{-1}(n_{ |\overleftarrow{\nabla}_y| } (t_y))^T \}^{\gamma_2\beta} 
\Bigr]\nonumber \\
&&~ + \theta (t_y-t_x)
\Bigl[ B^{-1}(n_{|\nabla_x|}(t_x))^{\alpha\gamma_1}\langle \theta|
 \bar{\phi}_{\xi,-}^{\gamma_2} (y) \phi_{\xi,+}^{\gamma_1}(x)|\theta\rangle
 B(n_{|\overleftarrow{\nabla}_y|} (t_y))^{\gamma_2 \beta} \nonumber \\
&&~~+\{\tau_3 B(n_{ |\nabla_x| }(t_x))^{T}\}^{\alpha\gamma_1}
\langle \theta| \bar{\phi}_{\xi,+}^{\gamma_2}(y) \phi_{\xi,-}^{\gamma_1}(y)
|\theta\rangle \{\tau_3 B^{-1}(n_{|\overleftarrow{\nabla}_y|} (t_y))^T \}^{\gamma_2\beta}
\Bigr],
\end{eqnarray}
where the fields, $\phi_{\xi,\pm}$ and $\bar{\phi}_{\xi,\pm}$, are 
defined in Eq.~(\ref{phixi-2}).

Due to the definition of the thermal vacuum (\ref{theta1}) and (\ref{theta2})
the thermal propagator, $D_0^{\alpha\beta}$, reduces to
\begin{eqnarray}
&& D_0^{\alpha\beta} (t_x, t_y,\Vec{x}-\Vec{y}) =
\Biggl[ B^{-1}(n_{|\nabla_x|}(t_x))^{\alpha\gamma_1} 
D_{0,R}^{\gamma_1 \gamma_2} (x-y) 
B(n_{|\overleftarrow{\nabla}_y|}(t_y))^{\gamma_2\beta}  \nonumber \\
&& +\{\tau_3 B(n_{|\nabla_x|}(t_x))^T \}^{\alpha\gamma_1} 
 D_{0,A}^{\gamma_1 \gamma_2} (x-y)
 \{ B^{-1}(n_{|\overleftarrow{\nabla}_y|}(t_y))^T\tau_3 \}^{\gamma_2 \beta}\Biggr],	
\label{ThermPro0}
\end{eqnarray}
where $ D_{0,R}^{\gamma_1 \gamma_2} (x-y)$ and 
$D_{0,A}^{\gamma_1 \gamma_2} (x-y)$ represent 
retarded and advanced parts of the propagator, respectively,
\begin{eqnarray}
&& D_{0,R}^{11} (x-y)=\int \frac{d^3\Vec{p}}{(2\pi)^3}
\theta(t_x-t_y) \frac{1}{2\omega_p} {\rm e}^{-i p \cdot (x-y)}, 
\label{ns_prop_com1} \\
&& D_{0,R}^{22} (x-y)=-\int \frac{d^3\Vec{p}}{(2\pi)^3}
\theta(t_y-t_x) \frac{1}{2\omega_p}{\rm e}^{-i p \cdot (x-y)}, 
\label{ns_prop_com2} \\
&& D_{0,A}^{11} (x-y)=\int \frac{d^3\Vec{p}}{(2\pi)^3}
\theta(t_y-t_x) \frac{1}{2\omega_p}{\rm e}^{i p\cdot (x-y)}, 
\label{ns_prop_com3} \\
&& D_{0,A}^{22} (x-y)=-\int \frac{d^3\Vec{p}}{(2\pi)^3}
\theta(t_x-t_y) \frac{1}{2\omega_p}{\rm e}^{i p \cdot (x-y)}, 	
\label{ns_prop_com4}	\\
&&{\rm other \ components }=0. \nonumber
\end{eqnarray}
Thus we reproduce the result obtained in Refs.~\citen{umezawa1}
and \citen{prop1}. The thermal propagator (\ref{ThermPro0}) 
has the same form as the one in an equilibrium system. The 
time dependence is introduced through the thermal Bogoliubov
transformation.

\section{Non-relativistic limit of the Boltzmann equation}

Here we take the non-relativistic limit of
the time evolution equation for the $\lambda\phi^4$ 
interaction model (\ref{BoltzEq2-loop}) and compare
it with the transport equation for the cold atom system.

The cold atom system is described by the Hamiltonian,
\begin{eqnarray}
H = \int d^3 \Vec{x} \left[ -\psi^{\dagger}(x) \frac{\nabla_x^2}{2m}\psi(x) 
+ \frac{g}{2} \psi^{\dagger}(x) \psi^{\dagger}(x) \psi(x) \psi(x) \right],
\label{mdl-nonrela1}
\end{eqnarray}
where $\psi$ represents a non-relativistic scalar field, 
\begin{eqnarray}
\psi (x) = \int \frac{d^3 \Vec{p}}{(2\pi)^3} a_p (t_x) e^{i\Vec{p}\cdot \Vec{x}} .
\end{eqnarray}
The quantum transport equation for this system is given by\cite{ColdAtm3}
\begin{eqnarray}
&&\dot{n}_p(t) = 4 g^2 {\rm Re} \int_{-\infty}^{t} ds 
\int \frac{d^3 \Vec{p}_1}{(2\pi)^3} \frac{d^3 \Vec{p}_2}{(2\pi)^3} 
\frac{d^3 \Vec{p}_3}{(2\pi)^3} \nonumber \\
&& \times (2\pi)^3 \delta^{(3)} (\Vec{p}_1 + \Vec{p}_2 - \Vec{p}_3 - \Vec{p}) 
e^{-i (\varepsilon_{p_1} + \varepsilon_{p_2} -\varepsilon_{p_3} - \varepsilon_{p})(t-s)}
\nonumber \\ 
&&\times \Bigr\{ n_{p_1}(s) n_{p_2}(s)(1+n_{p_3}(s))(1+n_p(s)) 
 - (1+n_{p_1}(s)) (1+n_{p_2}(s)) n_{p_3}(s) n_{p}(s) \Bigr\}, \nonumber \\
\label{BolzNR-1}
\end{eqnarray}
where $\varepsilon_p$ is the kinetic energy for the
non-relativistic field,
\begin{eqnarray}
\varepsilon_p = \frac{\Vec{p}^2}{2m}.
\end{eqnarray}

At the non-relativistic limit, $|\Vec{p}| \ll m$, 
the energy eigenvalue $\omega_p$ reduces to
\begin{eqnarray}
\omega_p  \approx m + \varepsilon_p. \label{nonR-engy}
\end{eqnarray}
We restrict the momentum for the scalar field
in $-p_\epsilon \le p_i \le p_\epsilon$ with a cut-off
parameter, $0 < p_\epsilon \ll m$. Thus the relativistic 
scalar field, $\phi_a$, is decomposed to be
\begin{eqnarray}
&&\phi_a (x) \approx \int_{-p_\epsilon}^{p_\epsilon} \frac{d^{3}\Vec{p}}{(2\pi)^3}
\frac{1}{\sqrt[]{2m}} \bigl\{ a_p(t_x) e^{i\Vec{p} \cdot \Vec{x}} 
+ a_p^\dagger (t_x) e^{-i \Vec{p}\cdot \Vec{x}} \bigr\} \nonumber \\
&&=\phi_{a,+}(x)+\phi_{a,-}(x),
\end{eqnarray}
where $\phi_{a,+}$ and $\phi_{a,-}$ indicate the positive and 
negative frequency parts.

The interaction for the cold atom system (\ref{mdl-nonrela1}) 
can be identified with the interaction,
$\phi_{a,+} \phi_{a,+} \phi_{a,-} \phi_{a,-}$. 
There are six corresponding terms in the interaction, $\lambda \phi_a^4$.
Thus we obtain the following correspondence between the
interaction terms for the cold atom system and the
$\lambda \phi_a^4$ model in the non-relativistic regime, 
\begin{eqnarray}
\frac{g}{2}\psi^\dagger(x) \psi^\dagger(x) \psi(x) \psi(x) 
=6\frac{\lambda}{4!} \phi_{a,+}(x) \phi_{a,+}(x) \phi_{a,-}(x) \phi_{a,-}(x),
\end{eqnarray}
We find that there is a correspondence if we make a replacement
\begin{eqnarray}
\lambda \leftrightarrow 8 m^2g. \label{nonRlLmd}
\end{eqnarray}
Since the transport equation (\ref{BolzNR-1}) comes from the
two-body scattering, we pick up terms which represent the 
two-body scattering in Eq.~(\ref{BoltzEq2-loop}). 
Hence Eq.~(\ref{BoltzEq2-loop}) is rewritten as
\begin{eqnarray}
&&\dot{n}_{p}(t_x) = 
\frac{\lambda^2}{2} 
 \int_{-\infty}^{t_x} dt_s \int \frac{d^3\Vec{k}_1}{(2{\pi})^3} \frac{d^3\Vec{k}_2}{(2{\pi})^3}
 \frac{d^3\Vec{k}_3}{(2{\pi})^3} 
	\frac{1}{8\omega_{p} \omega_{k_1} \omega_{k_2} \omega_{k_3} }	\nonumber \\
&&\times {\cos} \bigl\{ (\omega_{p}+\omega_{k_1}
-\omega_{k_2} - \omega_{k_3}) (t_x-t_s) \bigr\}	
 (2\pi)^3 \delta^{(3)}(\Vec{p} - \Vec{k}_1 - \Vec{k}_2 - \Vec{k}_3 )
 \nonumber \\
&&\times \Bigl[ n_{p}(t_s) n_{k_1}(t_s) (1+n_{k_2}(t_s)) (1 + n_{k_3}(t_s))
  - (1+n_p(t_s)) (1 + n_{k_1}(t_s)) n_{k_2}(t_s) n_{k_3}(t_s) \Bigr]. \nonumber \\
	\label{BolzNR-2}
\end{eqnarray}
Substituting (\ref{nonR-engy}) and (\ref{nonRlLmd}) into Eq.~(\ref{BolzNR-2})
we obtain the quantum Boltzmann equation in the non-relativistic regime,
\begin{eqnarray}
&&\dot{n}_{p}(t_x) = 4 g^2 
 \int_{-\infty}^{t_x} dt_s \int \frac{d^3\Vec{k}_1}{(2{\pi})^3} \frac{d^3\Vec{k}_2}{(2{\pi})^3}
 \frac{d^3\Vec{k}_3}{(2{\pi})^3} \nonumber \\
&&\times {\cos} \bigl\{ (\varepsilon_{p} + \varepsilon_{k_1}
 - \varepsilon_{k_2} - \varepsilon_{k_3}) (t_x-t_s) \bigr\}	
 (2\pi)^3 \delta^{(3)}(\Vec{p} - \Vec{k}_1 - \Vec{k}_2 - \Vec{k}_3 )
 \nonumber \\
&&\times \Bigl[ n_{p}(t_s) n_{k_1}(t_s) (1+n_{k_2}(t_s)) (1 + n_{k_3}(t_s))
  - (1+n_p(t_s)) (1 + n_{k_1}(t_s)) n_{k_2}(t_s) n_{k_3}(t_s) \Bigr]. \nonumber \\
	\label{BolzNR-3}
\end{eqnarray}
In homogeneous system the quantum Boltzmann equation (\ref{BolzNR-3}) and 
Eq.~(\ref{BolzNR-1}) become identical. Therefore the time evolution 
equation (\ref{BoltzEq2-loop}) is consistent with the quantum transport
equation (\ref{BolzNR-1}).

\end{document}